\documentclass[preprint,12pt]{elsarticle}
\usepackage[breaklinks=true]{hyperref}
\usepackage{soul,xcolor,bm}
\usepackage{geometry}
\usepackage{multirow}
\biboptions{numbers,sort&compress}

\usepackage{amssymb}
\journal{arXiv}

\begin{document}

\begin{frontmatter}



\title{A Simple Model for Short-Range Ordering Kinetics in Multi-Principal Element Alloys}


\author[inst1,inst5]{Anas Abu-Odeh}

\affiliation[inst1]{organization={Materials Science and Engineering Division, National Institute of Standards and Technology},%Department and Organization
            city={Gaithersburg},
            postcode={20899}, 
            state={MD},
            country={USA}}

\affiliation[inst5]{organization={Department of Materials Science and Engineering, University of California},%Department and Organization
            city={Berkeley},
            postcode={94720}, 
            state={CA},
            country={USA}}

\author[inst23]{Bin Xing}
\author[inst23,inst3]{Penghui Cao}

\affiliation[inst23]{organization={Department of Materials Science and Engineering, University of California},%Department and Organization
            city={Irvine},
            postcode={92697}, 
            state={CA},
            country={USA}}
            
\affiliation[inst3]{organization={Department of Mechanical and Aerospace Engineering, University of California},%Department and Organization
            city={Irvine},
            postcode={92697}, 
            state={CA},
            country={USA}}

\author[inst4]{Blas Pedro Uberuaga}

\affiliation[inst4]{organization={Materials Science and Technology Division, Los Alamos National Laboratory},%Department and Organization
            city={Los Alamos},
            postcode={87545}, 
            state={NM},
            country={USA}}

\author[inst5,inst6]{Mark Asta}

\affiliation[inst6]{organization={Material Sciences Division, Lawrence Berkeley National Laboratory},%Department and Organization
            city={Berkeley},
            postcode={94720}, 
            state={CA},
            country={USA}}

\begin{abstract}
Short-range ordering (SRO) in multi-principal element alloys influences material properties such as strength and corrosion. While some degree of SRO is expected at equilibrium, predicting the kinetics of its formation is challenging. We present a simplified isothermal concentration-wave (CW) model to estimate an effective relaxation time of SRO formation. Estimates from the CW model agree to within a factor of five with relaxation times obtained from kinetic Monte Carlo (kMC) simulations when above the highest ordering instability temperature. The advantage of the CW model is that it only requires mobility and thermodynamic parameters, which are readily obtained from alloy mobility databases and Metropolis Monte Carlo simulations, respectively. The simple parameterization of the CW model and its analytical nature make it an attractive tool for the design of processing conditions to promote or suppress SRO in multicomponent alloys.
\end{abstract}

%\begin{keyword}
%keyword one \sep keyword two
%\end{keyword}

\end{frontmatter}


Short-range ordering (SRO) represents non-random correlations between atoms of different species in solid solution alloys. It influences properties such as strength \cite{Antillon2020,Antillon2021,Cao2022} and corrosion \cite{Liu2018,Xie2021} of multicomponent and multiprincipal element alloys. Predicting the equilibrium level of SRO is a prerequisite to predicting its effect on material properties. Monte Carlo simulations enabled through accurate descriptions of interatomic interactions are computationally assessable for this purpose \cite{ferrari2023simulating}. However, the kinetics of SRO formation are less easily predicted. While kinetic Monte Carlo (kMC) simulations provide an accurate computational framework, they are reliant on an extensive number of vacancy hopping energy barriers to enable the parameterization of their local-environment dependencies. In this paper, we use an analytical concentration wave (CW) model to predict SRO kinetics in a multicomponent solid solution alloy. This approach has been shown to successfully predict SRO kinetics in binary alloys \cite{Abu-Odeh2023}. The CW model circumvents the need to explicitly describe vacancy hopping by combining thermodynamic information from atomistic simulations and mobility information, which in principle can be obtained from experimentally-assessed kinetic databases \cite{Campbell2009}.

Here we will briefly review the necessary aspects of the CW model of multicomponent SRO kinetics as outlined in Ref. \cite{Fontaine1973} for cubic systems. {The CW model assumes that the inital state of the system is not too far away from equilibrium.} We consider a $n$-component substitutional system where one of the species is a vacancy and the other ($n-1$) correspond to the different atomic components. We consider the time ($t$) dependence of the Fourier transform of the $(n-1)\times(n-1)$ matrix of pair correlation functions $\textbf{Q}(\vec{k},t)$, with wave-vector ($\vec{k}$). The matrix elements $Q_{ij}(\vec{k},t) = \Delta c_i(\vec{k},t) \Delta c_j^{*}(\vec{k},t)$ are defined in terms of the amplitudes of the concentration waves:

\begin{equation} \label{eq:eqconc}
    \Delta c_i(\vec{k},t) = \frac{1}{N} \sum_{p=1}^N [c_i(\vec{r}_p,t)-\bar{c}_i]\mathrm{e}^{-2\pi \vec{k} \cdot \vec{r}_p}
\end{equation}
where $c_i(\vec{r}_p$,t) takes a value of one (zero) if species $i$ is present (absent) at the lattice site position $\vec{r_p}$, and $\bar{c}_i$ denotes the average site fraction of species $i$.

In the CW model, the time evolution of the eigenvalues of $\textbf{Q}(\vec{k},t)$ is given by:

\begin{equation} \label{eq:eqeig}
    Q_i (\vec{k},t) = [Q_i (\vec{k},0)-Q_i (\vec{k},\infty)]\mathrm{exp}[-t/\tau_i(\vec{k})]+Q_i (\vec{k},\infty)
\end{equation}
where $Q_i(\vec{k},0)$ and $Q_i(\vec{k},\infty)$ are initial and equilibrium values, respectively, and $\tau_i(\vec{k})$ is a characteristic relaxation time.  The value of $\tau_i(\vec{k})$ is related to an eigenvalue of the product of a $(n-1)\times(n-1)$ mobility matrix ($\textbf{M}$) and a thermodynamic-factor matrix ($\mathbf{\Psi}(\vec{k})$){, which is the Hessian matrix of the free energy of the solid-solution with respect to amplitudes of different concentration waves at $\vec{k}$. The product matrix, ($\textbf{M}\mathbf{\Psi}(\vec{k})$), is a $\vec{k}$-vector dependent diffusion matrix, which becomes the continuum diffusion matrix in the long-wavelength limit.}

Within this model, the value of the real-space pair-correlations, which are related to the Warren-Cowley (WC) parameters \cite{DeFontaine1971}, at any time are obtained through transforming the $n-1$ values of $Q_i(\vec{k},t)$ into a pair-correlation matrix in the original concentration space, which must be done for every $\vec{k}$-point, and carrying out an inverse Fourier transformation. To simplify the approach, we will adopt some assumptions which are shown below to lead to accurate predictions.
%For a $n$-component system, there are $n-1$ evolution equations as a function of time, $t$, for a given $\vec{k}$-point in reciprocal space given by:

%\begin{equation} \label{eq:1}
%    Q_i (\vec{k},t) = [Q_i (\vec{k},0)-Q_i (\vec{k},\infty)]\mathrm{exp}[-t/\tau_i(\vec{k})]+Q_i (\vec{k},\infty)
%\end{equation}
%where $Q_i (\vec{k},\infty)$ is an eigenvalue of a ($n-1$)$\times$($n-1$) equilibrium pair-correlation matrix in reciprocal space, $Q_i (\vec{k},t)$ is the transient value of $Q_i(\vec{k})$, $Q_i (\vec{k},0)$ is the initial value of $Q_i(\vec{k})$, and $\tau_i (\vec{k})$ is a characteristic relaxation time for $Q_i (\vec{k})$. $\tau_i (\vec{k})$ is related to an eigenvalue of the product of a ($n-1$)$\times$($n-1$) mobility ($\textbf{M}$) and a ($n-1$)$\times$($n-1$) thermodynamic matrix ($\bm{\Psi}(\vec{k})$). Within this model, the value of the real-space pair-correlations, which are related to the WC parameters \cite{DeFontaine1971}, at any time $t$ are obtained through transforming the $n-1$ values of $Q_i (\vec{k},t)$ into a pair-correlation matrix in the original concentration space, which must be done for every $\vec{k}$-point, and carrying out an inverse Fourier transformation for the entire Brillouin zone. As such an approach is cumbersome, we will adopt some simplifying assumptions.

The first simplifying assumption is that the SRO kinetics are controlled by a dominant $\vec{k}$-point, $\vec{k}_D$. This can be characterized as the $\vec{k}$-point that has the smallest determinant of $\bm{\Psi}(\vec{k})$ \cite{Fontaine1979}. {A small value of the determinant, which is equal to the product of the eigenvalues of $\bm{\Psi}(\vec{k})$, means that there is a low thermodynamic cost associated with the formation of a characteristic concentration wave at $\vec{k}$. This causes the solid solution to be more susceptible to the formation of pair-correlations associated with $\vec{k}$.} The second simplifying assumption is that at $\vec{k}_D$, SRO kinetics are limited by the slowest characteristic relaxation time, $\tau_s (\vec{k_D})$. The goal of this study is to present a procedure for calculating $\tau_s (\vec{k_D})$ and compare this characteristic SRO relaxation time to kMC simulation results. We treat kMC simulations results as a ground truth due to lack of readily available experimental results for SRO kinetics.

For a given composition and temperature, the values of $\textbf{M}$, $\bm{\Psi}(\vec{k}_D)$, and the lattice constant, $a$, allow for the calculation of $\tau_s (\vec{k_D})$ through:

\begin{equation} \label{eq:2}
    \tau_s^{-1} (\vec{k_D}) = 2\beta(\vec{k}_D)\lambda_s(\vec{k}_D)
\end{equation}
where $\beta(\vec{k}_D)$ represents a lattice sum over vectors representing the {$Z$} nearest-neighbor positions ($\vec{r}_{nn}$) of an atom \footnote{We note that in Eq. 8 in \cite{Abu-Odeh2023} the factor of $2\pi$ is absorbed into the definition of the wave-vector and is not written in front of $\vec{k}$. The factor of $1/a^2$ is omitted in \cite{Abu-Odeh2023} as the lattice constant was set to 1 in the kMC simulations.}:

\begin{equation} \label{eq:3}
    \beta(\vec{k}_D) = \frac{1}{a^2} \sum_{nn = 1}^{Z} [1-\mathrm{cos}(2\pi \vec{k}_D \cdot \vec{r}_{nn})]
\end{equation}
and $\lambda_s(\vec{k}_D)$ is the smallest eigenvalue of the matrix product of $\textbf{M}$ and $\bm{\Psi}(\vec{k}_D)$.

We adopt the common practice of treating $\textbf{M}$ as a diagonal matrix \cite{Agren2012}, where its elements are related to tracer diffusion coefficients ($D^*_i$) \cite{Andersson1992}. Thus, we can parameterize $\textbf{M}$ while treating vacancies as a conserved species through:

\begin{equation} \label{eq:5}
    M_{ii} = \frac{D_i^* \bar{c}_i}{k_B T}
\end{equation}
where $k_B$ is the Boltzmann constant, and $T$ is the temperature. This results in a ($n-1$)$\times$($n-1$) mobility matrix where the vacancies are the $n^{th}$ species. For real alloy systems, the tracer diffusion coefficients can be obtained from mobility databases such as those used in the DICTRA\footnote{Certain commercial equipment, instruments, or materials are identified in this paper in order to specify the experimental procedure or concept adequately. Such identification is not intended to imply recommendation or endorsement by the National Institute of Standards and Technology, nor is it intended to imply that the materials or equipment identified are necessarily the best available for the purpose.} software package \cite{Borgenstam2000}.

$\bm{\Psi}(\vec{k}_D)$ is obtained through the following equation \cite{Fontaine1973}:

\begin{equation} \label{eq:6}
    \textbf{Q}(\vec{k}_D,\infty) = \frac{k_B T}{N} \bm{\Psi}^{-1}(\vec{k}_D)
\end{equation}
where $N$ is the total number of atoms in the system and $\bm{\Psi}^{-1}(\vec{k}_D)$ is the inverse of $\bm{\Psi}(\vec{k}_D)$. The elements of $\textbf{Q}(\vec{k}_D,\infty)$ are given as the equilibrium ensemble average values:

\begin{equation} \label{eq:7}
    Q_{ij}(\vec{k}_D,\infty) = \langle \Delta c_i(\vec{k}_D,\infty) \Delta c_j^*(\vec{k}_D,\infty)\rangle.
\end{equation}
Eqs. \ref{eq:eqconc}, \ref{eq:6}, and \ref{eq:7} allow for the calculation of $\bm{\Psi}(\vec{k}_D)$ through equilibrium pair-correlations obtained from Monte Carlo simulations in the canonical ensemble. Once $\textbf{M}$ and $\bm{\Psi}(\vec{k}_D)$ are determined, $\tau_s(\vec{k}_D)$ is obtained through Eq. \ref{eq:2}.

To test the validity of the CW model, we will compare it to results from kMC simulations of an equiatomic body-centered cubic NbMoTa alloy \cite{kmcData}. A neural-network kMC model was trained on approximately $7\times10^5$ vacancy hopping energy barriers for various compositions of the Nb-Mo-Ta system as evaluated from a machine-learned interatomic potential \cite{Yin2021}. Further details are found in Ref. \cite{Xing2023}. Supercells of 1024 atoms  with 341 Nb atoms, 341 Mo atoms, 341 Ta atoms, and 1 vacancy were used for most kMC simulations. Tracer diffusion coefficients in the equilibrium SRO state ($D^*_{eq}$) were obtained from measuring the mean-squared displacement of different species as a function of time. Tracer diffusion coefficients in the random state ($D^*_{rand}$) were calculated using a supercell size of 128,000 atoms with 1 vacancy and scaled to a vacancy concentration of 1/1024. These allow for comparisons of the mobility matrix as given by Eq. \ref{eq:5} as a function of these different states, as this system is known to have its tracer diffusion coefficients reduced in the presence of SRO \cite{Xing2022}. 

As NbMoTa shows a clear B2 ordering of Mo and Ta atoms at lower temperatures \cite{Xing2023}, $\vec{k}_D$ represents the $\frac{1}{a}\langle 100 \rangle$ wave-vector in this system \cite{Sluitter1988}. Instead of canonical Monte Carlo simulations, we use previous long-time kMC simulations \cite{Xing2023} of this system consisting of $2\times10^7$ kMC hops to sample equilibrium states. We use a total of 800 snapshots from every $2\times10^4$ kMC hops after the initial $4\times10^6$ (long enough for all states at temperatures above {850 K} to reach equilibrium \cite{Xing2023}) to obtain an average of the pair-correlations using Eq. \ref{eq:7}. {We note that if Metropolis Monte Carlo data using the same model was available, it would provide the same information.} These pair-correlations are used to determine $\bm{\Psi}(\vec{k}_D)$ as well as the instability temperature, {below which the solid solution phase can reduce its free energy by spontaneously undergoing a long-range order transformation.}

\begin{figure}[h!]
    \centering
    \includegraphics[width=1\textwidth]{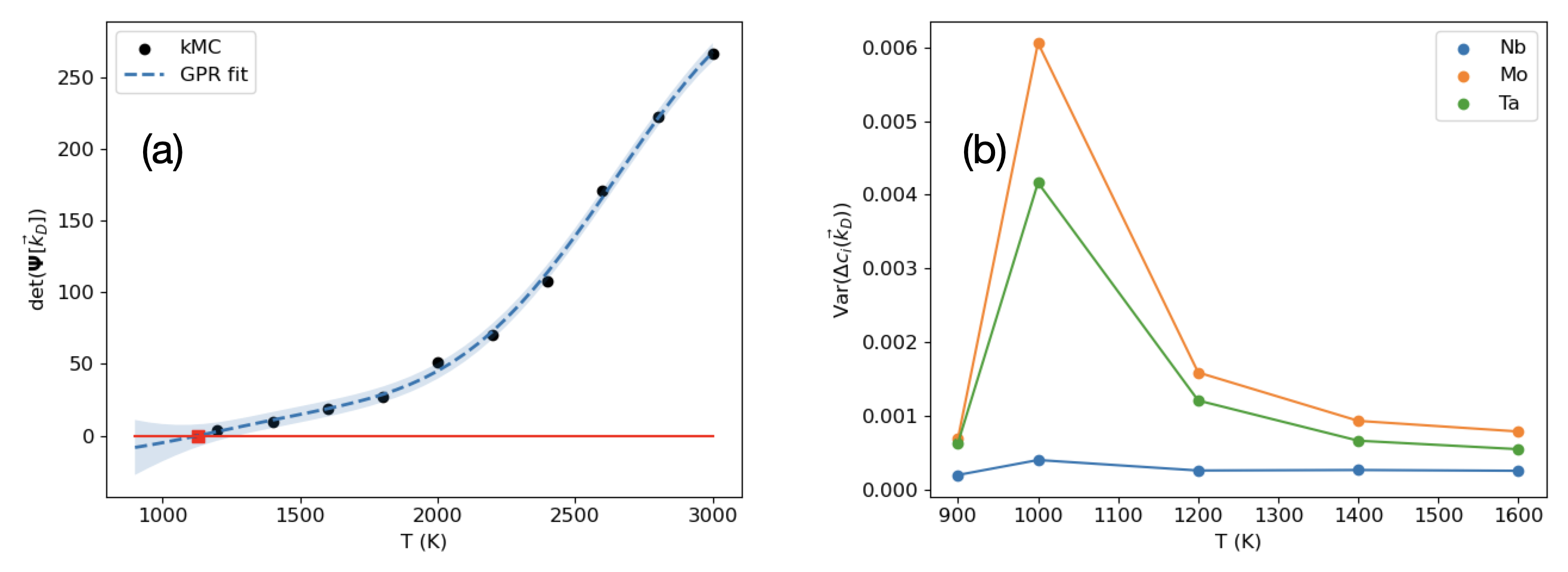}
    \caption{(a) Determinant of the thermodynamic matrix {(units are in eV$^3$)} at the $\frac{1}{a}\langle 100 \rangle$ wave-vector as a function of temperature. The points represent data directly obtained from long-time kMC simulations, while the dashed blue line represents a fit to a Gaussian process regression (GPR) model. The shaded region represents the estimated uncertainty of the model. The red square is the intersection of the GPR model with the value of zero at a temperature of 1128 K. (b) Variance of the CW amplitudes for different species obtained from long-time kMC simulations as a function of temperature. Lines are for guiding the eye.}
    \label{fig:instability}
\end{figure}

Figure \ref{fig:instability} shows an analysis of the stability of the disordered solid solution. In Figure \ref{fig:instability}(a), data points obtained from long-time kMC simulations represent the determinant of the $\bm{\Psi}(\vec{k}_D)$ matrix as a function of temperature. The solid solution is unstable with respect to a CW with a wave-vector of $\vec{k}_D$ when this determinant reaches zero, {which means that one of the eigenvalues of $\bm{\Psi}(\vec{k}_D)$ has also reached zero and that the amplitude of a characteristic concentration wave can grow without any thermodynamic penalty}. Using a Guassian process regression (GPR) model (implemented in the \textit{scikit-learn} package \cite{Pedregosa2011}) fit to the kMC data, we estimate the instability temperature to be 1128 K. Due to the uncertainty of the GPR model at lower temperatures, we also calculate the temperature-dependent variance of $\Delta c_i (\vec{k}_D)$ for Nb, Mo, and Ta in Figure \ref{fig:instability}(b). A spike is observed around 1000 K. Therefore, we estimate the instability temperature for the disordered state of our system to be between 1000 K and 1128 K.

With $\textbf{M}$ and $\bm{\Psi}(\vec{k}_D)$ determined, we obtain the CW model estimate of $\tau_s({\vec{k}_D})$ from Eq. \ref{eq:2}. We compare this value to relaxation times obtained from shorter kMC simulations of SRO evolution starting from a random state. Five independent kMC simulations with different initial random configurations were carried out at temperatures of 1200 K and 2600 K for $2\times10^5$ kMC hops each. These temperatures were chosen to assess the level of prediction of the CW model at temperatures both near and far from the estimated instability temperature range. We note that the isothermal SRO relaxation kinetics in this study are not representative of the kinetics of the true NbMoTa system due to the artificially high vacancy concentration of 1/1024. {A more realistic relaxation time could be orders of magnitude slower.} This does not limit the applicability of the CW model to real systems, as mobilities obtained from kinetic databases are representative of diffusion with an equilibrium concentration of vacancies. Of course, these databases do not describe mobilities when a non-equilibrium concentration of point defects are present, such as during rapid quenching or when under irradiation.

\begin{figure}[h!]
    \centering
    \includegraphics[width=1\textwidth]{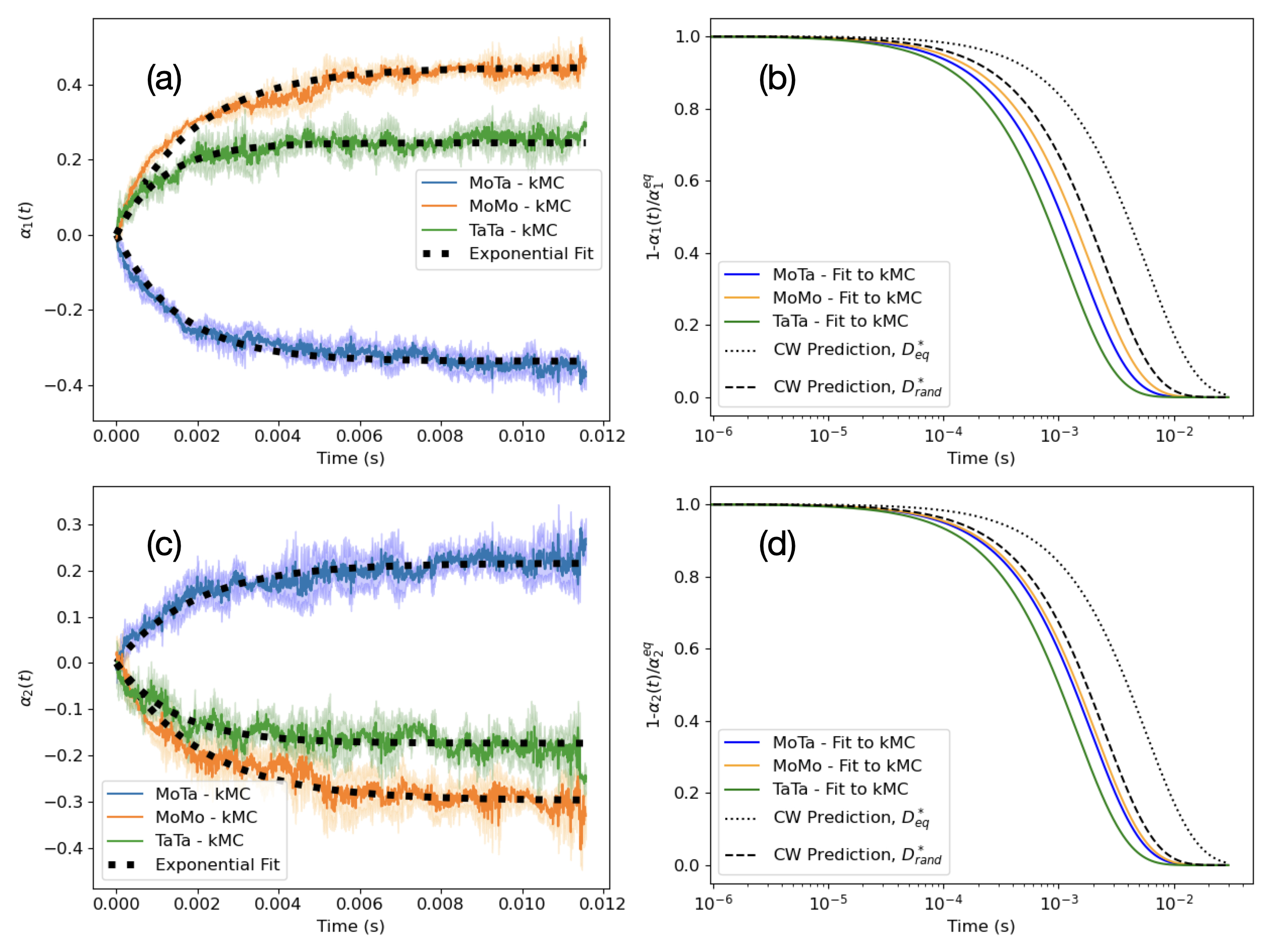}
    \caption{(a) Bin-averaged time dependent WC parameters for the first nearest neighbor shells between MoTa, MoMo, and TaTa pairs obtained from five independent kMC simulations at 1200 K. The shaded regions represent +/- one standard deviation of each bin. The black dotted lines represent exponential fits to the binned data. (b) Comparisons of exponential kinetics as fit to kMC data for different pairs within a first nearest neighbor shell as well as to CW predictions using tracer diffusivities obtained from an equilibrium SRO state ($D^*_{eq}$) or a random state ($D^*_{rand}$). (c) and (d) are the same as (a) and (b), respectively, except as evaluated for the second nearest neighbor shell.}
    \label{fig:1200_cw}
\end{figure}

Figures \ref{fig:1200_cw}(a) and (c) show the kinetics of the WC parameters from kMC simulations for the first and second nearest neighbor shells at 1200 K. The WC parameter between species $i$ and $j$ for the $m^{th}$ shell is evaluated as \cite{Walsh2023}:

\begin{equation} \label{eq:9}
    \alpha_{m}^{ij} = 1 - \frac{P_m (i|j)}{\bar{c}_i}
\end{equation}
where $P_m (i|j)$ is the probability of finding atom species $i$ around the $m^{th}$ shell of a $j$ atom. We only show three of the possible different pairs in the NbMoTa system as that is the number of linearly independent pair-correlations in a three component system (ignoring the vacancy) \cite{DeFontaine1971}. The time-dependence of the WC parameters can be reliably fit to an exponential kinetics model:

\begin{equation} \label{eq:10}
    1-\frac{\alpha_{m}^{ij}(t)}{\alpha_{m}^{ij,eq}} = \mathrm{exp}[-t/\tau_m^{ij}]
\end{equation}
where $\alpha_{m}^{ij,eq}$ is the WC parameter at equilibrium (extracted from the long-time kMC simulations), and $\tau_m^{ij}$ is a relaxation time that is obtained from fitting to the kMC results. 

In Figures \ref{fig:1200_cw}(b) and (d) we plot the behavior of the exponential model with relaxation times obtained from fitting to kMC results, as well as with relaxation times obtained from the CW model using Eq. \ref{eq:2}. While the relaxation times obtained from kMC results have a slight dependence on the type of pairs, they are well within the same order of magnitude of each other. Regardless of whether $D^*_{eq}$ or $D^*_{rand}$ coefficients are used to parameterize $\textbf{M}$, the kinetics from the CW model are within the same order of magnitude as the exponential model fit to kMC results. However, the CW model using $D^*_{rand}$ results in a better prediction. As our simulations start with a system closer to random and evolves toward the SRO state, the vacancy kinetics are at least initially better described by the random state. However, at worst the CW predictions using $D^*_{eq}$ overestimates the relaxation time only by a factor of five. This is encouraging as tracer diffusivities obtained from mobility databases will describe diffusion in an equilibrium state. The first three columns of Table \ref{tab:1200_ratio} lists the ratios between the CW relaxation times and those extracted from kMC at a temperature of 1200 K.

\begin{table}[h!]
    \centering
	\begin{tabular}{l | l l l | l l l}
        \multirow{2}{*}{} &  & 1200 K & & & 2600 K & 
        \\ & MoTa & MoMo & TaTa & MoTa & MoMo & TaTa\\ \hline &&&&&& \\
		$\tau_s(\vec{k}_D)/\tau^{ij}_1$ ($D^*_{eq}$) & 3.7 & 3.0 & 5.0 & 0.7 & 0.7 & 0.8 \\ &&&&&& \\
		$\tau_s(\vec{k}_D)/\tau^{ij}_2$ ($D^*_{eq}$) & 3.0 & 2.7 & 3.9 & 1.6 & 1.8 & 1.4 \\ &&&&&& \\
		$\tau_s(\vec{k}_D)/\tau^{ij}_1$ ($D^*_{rand}$) & 1.6 & 1.3 & 2.2 & 0.7 & 0.7 & 0.7 \\ &&&&&& \\
		$\tau_s(\vec{k}_D)/\tau^{ij}_2$ ($D^*_{rand}$) & 1.3 & 1.2 & 1.7 & 1.5 & 1.7 & 1.3 \\ &&&&&& \\

    \end{tabular}
	\caption{\label{tab:1200_ratio} Ratio between relaxation times obtained from Eq. \ref{eq:2} to relaxation times obtained from kMC using Eq. \ref{eq:10} as a function of pairs, nearest neighbor shells, and choices of tracer diffusivities at temperatures of 1200 K and 2600 K.}
\end{table}

\begin{figure}[h!]
    \centering
    \includegraphics[width=1\textwidth]{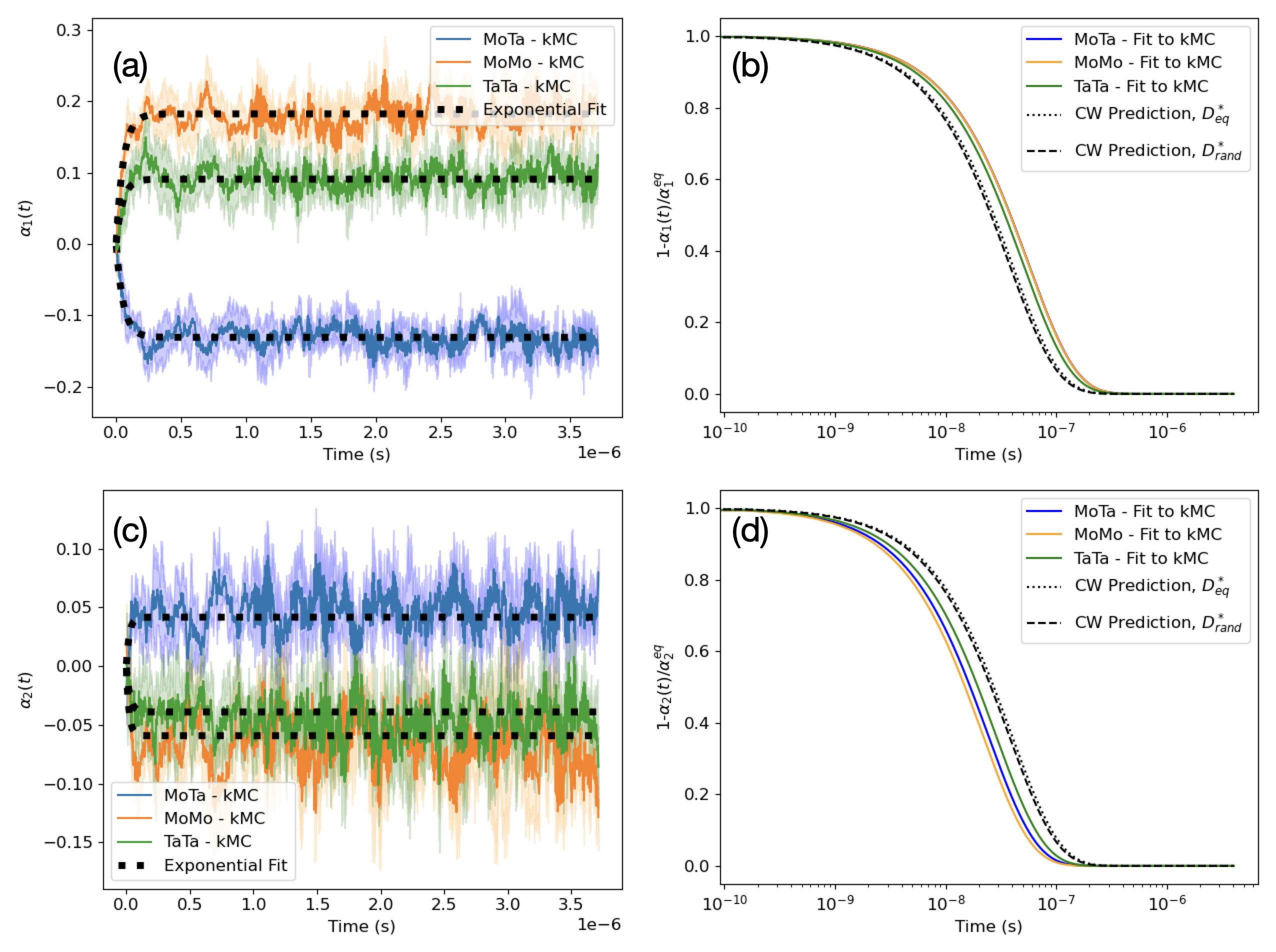}
    \caption{(a) Bin-averaged time dependent WC parameters for the first nearest neighbor shells between MoTa, MoMo, and TaTa pairs obtained from five independent kMC simulations at 2600 K. The shaded regions represent +/- one standard deviation of each bin. The black dotted lines represent exponential fits to the binned data. (b) Comparisons of exponential kinetics as fit to kMC data for different pairs within a first nearest neighbor shell as well as to CW predictions using tracer diffusivities obtained from an equilibrium SRO state ($D^*_{eq}$) or a random state ($D^*_{rand}$). (c) and (d) are the same as (a) and (b), respectively, except as evaluated for the second nearest neighbor shell.}
    \label{fig:2600_cw}
\end{figure}

Figures \ref{fig:2600_cw}(a) and (c) show the kinetics of the WC SRO parameters from kMC simulations for the first and second nearest neighbor shells at 2600 K. Similar to the 1200 K case, we find that the kMC results are well fit to Eq. \ref{eq:10}. Figures \ref{fig:2600_cw}(b) and (d) show that the effective relaxation time obtained from the kMC results are less dependent on the types of pairs at a temperature of 2600 K than at a temperature of 1200 K. We find that the difference between the CW prediction when using $D^*_{rand}$ or $D^*_{eq}$ is minimal. At higher temperatures, SRO is weaker, so its effect on vacancy diffusion behavior diminishes. Additionally, for the first nearest neighbor shell  the CW predictions seem to slightly underestimate the relaxation time, while for the second nearest neighbor shell the CW predictions seem to slightly overestimate the relaxation time. Regardless, the CW model provides a description of SRO kinetics that agrees well with the time scale obtained from kMC simulations at the temperature of 2600 K, with the worst disagreement being approximately a factor of two. The last three columns of Table \ref{tab:1200_ratio} lists the ratios between the CW relaxation times and those extracted from kMC at a temperature of 2600 K. Ultimately, the slowest characteristic relaxation time predicted from the CW model agrees with the SRO relaxation times obtained from kMC simulations to within an order of magnitude at and above temperatures of approximately 1.1 to 1.2 times the instability temperature.

{For the sake of comparison, we also look at the slowest relaxation time at two other $\vec{k}$-vectors: $\frac{1}{2a}\langle 110 \rangle$ and $\frac{1}{2a}\langle 111 \rangle$. These represent two other special points for the BCC lattice \cite{Sluitter1988}. The relaxation times using $D^*_{eq}$ at the temperatures of both 1200 K and 2600 K are listed in Table \ref{tab:other_ks}, along with the determinant of the $\bm{\Psi}(\vec{k})$ matrix and the value of $\beta(\vec{k})$ for each $\vec{k}$-vector. At 1200 K, the slowest relaxation time is associated with $\vec{k}_D$ along with the smallest determinant of $\bm{\Psi}(\vec{k})$. This is in agreement with our first assumption. At 2600 K, $\vec{k}_D$ still has the smallest determinant of $\bm{\Psi}(\vec{k})$, but no longer has the slowest relaxation time. At this temperature, all three relaxation times are relatively close to each other. This is likely due to the fact that the temperature-independent lattice sum, $\beta(\vec{k})$, is twice as large at $\vec{k}_D$ than the other two $\vec{k}$-vectors. While this shows that our first assumption is not strictly true at higher temperatures, the relaxation time obtained at $\vec{k}_D$ is still a useful estimate as shown earlier.}

\begin{table}[h!]
    \centering
	\begin{tabular}{l | l l| l l| l}
        \multirow{2}{*}{} & 1200 K & & 2600 K & &
        \\ $\vec{k}$ & $\tau_s({\vec{k}})$ & det($\bm{\Psi}(\vec{k})$) & $\tau_s({\vec{k}})$ & det($\bm{\Psi}(\vec{k})$) & $\beta(\vec{k})$ \\ \hline &&&&& \\
		$\frac{1}{a}\langle 100 \rangle$ ($\vec{k}_D$) & 5.7$\times$10$^{-3}$ s & 4 eV$^3$ & 3.9$\times$10$^{-8}$ s & 171 eV$^3$& 16  \\ &&&&& \\
		$\frac{1}{2a}\langle 110 \rangle$ & 0.9$\times$10$^{-3}$ s & 51 eV$^3$& 3.6$\times$10$^{-8}$ s & 397 eV$^3$& 8 \\ &&&&& \\
		$\frac{1}{2a}\langle 111 \rangle$ & 1.5$\times$10$^{-3}$ s & 35 eV$^3$& 4.5$\times$10$^{-8}$ s & 333 eV$^3$& 8 \\ &&&&& \\

    \end{tabular}
	\caption{\label{tab:other_ks}  The slowest relaxation time at different $\vec{k}$-vectors, along with the determinant of  $\bm{\Psi}(\vec{k})$ at the temperatures of 1200 K and 2600 K. The last column shows the values of $\beta(\vec{k})$.}
\end{table}

One inconvenience associated with the above approach is that the calculation of the thermodynamic matrix through Eq. \ref{eq:6} requires Monte Carlo simulations that include at least one vacancy, which requires the parameterization of vacancy-atom interactions \cite{VanderVen2005}. For systems without a vacancy, the number of rows and columns in the $\bm{\Psi}(\vec{k}_D)$ matrix decrease by one each, and the matrix product of $\textbf{M}$ with $\bm{\Psi}(\vec{k}_D)$ cannot be performed when assuming a conserved network of lattice sites. It is possible to reduce the size of the $\textbf{M}$ matrix by assuming that a sufficient density of sources and sinks exist to instantaneously maintain a local equilibrium of vacancy concentration \cite{Andersson1992}. However such an assumption does not seem warrented as, firstly, we have no sources or sinks in the kMC simulations we are comparing to, and, secondly, in real material systems the characteristic length scale for SRO (well away from the instability temperature) is on the order of the lattice spacing while the characteristic length scale of microstructural sources and sinks is typically much larger \cite{Nastar2014}. To avoid dealing with vacancies in Monte Carlo, we follow Ref. \cite{Fontaine1979} and adopt a parameterization of the matrix elements of $\bm{\Psi}(\vec{k}_D)$ through:

\begin{equation} \label{eq:11}
    \Psi_{ii}(\vec{k}_D) = -2f_{in}(\vec{k}_D) + k_B T \Bigl(\frac{1}{\bar{c}_n}+\frac{1}{\bar{c}_i}\Bigr)
\end{equation}

\begin{equation} \label{eq:12}
    \Psi_{ij}(\vec{k}_D) = f_{ij}(\vec{k}_D) - f_{in}(\vec{k}_D) - f_{nj}(\vec{k}_D) + \frac{k_B T}{\bar{c}_n}
\end{equation}
where the concentration dependent terms represent the thermodynamically ideal contributions to $\bm{\Psi}(\vec{k}_D)$, and the $f_{ij}(\vec{k}_D)$ terms represent the thermodynamically non-ideal contributions associated with species $i$ and $j$. In treating the vacancy, the $n^{th}$ species, as ideal, all the non-ideal contributions with the subscript of $n$ are set to zero, and $\bar{c}_n$ is set to some arbitrarily dilute concentration, with the other concentrations adjusted to maintain a sum of one. Using Eqs. \ref{eq:11} and \ref{eq:12} requires the non-ideal contributions of non-vacancy species. This is obtained by fitting Eqs. \ref{eq:11} and \ref{eq:12} to the ($n^*-1$)$\times$($n^*-1$) $\bm{\Psi}^*(\vec{k}_D)$ matrix (where $n^* = n-1$) determined from Eq. \ref{eq:6} using canonical Monte Carlo simulations without a vacancy. Assuming that a dilute concentration of vacancies will not affect the non-ideal terms obtained from this matrix, they can be used to construct a ($n-1$)$\times$($n-1$) $\bm{\Psi}(\vec{k}_D)$ matrix with Eqs. \ref{eq:11} and \ref{eq:12} while treating the vacancy as the $n^{th}$ species with ideal interactions.

To evaluate how well this approach works, we once again calculate the pair-correlations from Eq. \ref{eq:7} from the 800 snapshots obtained from long-time kMC, but this time we replace the vacancy in each snapshot with a Ta atom. We note that the following results are nearly identical when replacing the vacancy with either a Nb or Mo atom instead, which is to be expected given the low vacancy concentration of 1/1024 in the simulated system. We obtain a ($n^*-1$)$\times$($n^*-1$) $\bm{\Psi}^*(\vec{k}_D)$ matrix using Eq. \ref{eq:6}, and fit the non-ideal contributions in Eqs. \ref{eq:11} and \ref{eq:12} to this matrix. Those equations are then used to approximate the ($n-1$)$\times$($n-1$) $\bm{\Psi}(\vec{k}_D)$ matrix with an ideal vacancy with an arbitrarily chosen value of the vacancy concentration, $c_V$ (which is also $\bar{c}_n$). This is used to approximate all the characteristic relaxation times {at $\vec{k}_D$} from the CW model, not just the slowest one, through equations analogous to Eq. \ref{eq:2} for the faster relaxation times. We emphasize that $c_V$ is not a true vacancy concentration, just a value chosen to approximate the dilute ideal contribution to $\bm{\Psi}(\vec{k}_D)$. The choice of $c_V$ does not affect $\textbf{M}$, which is dependent on the true vacancy concentration.

\begin{figure}[h!]
    \centering
    \includegraphics[width=1\textwidth]{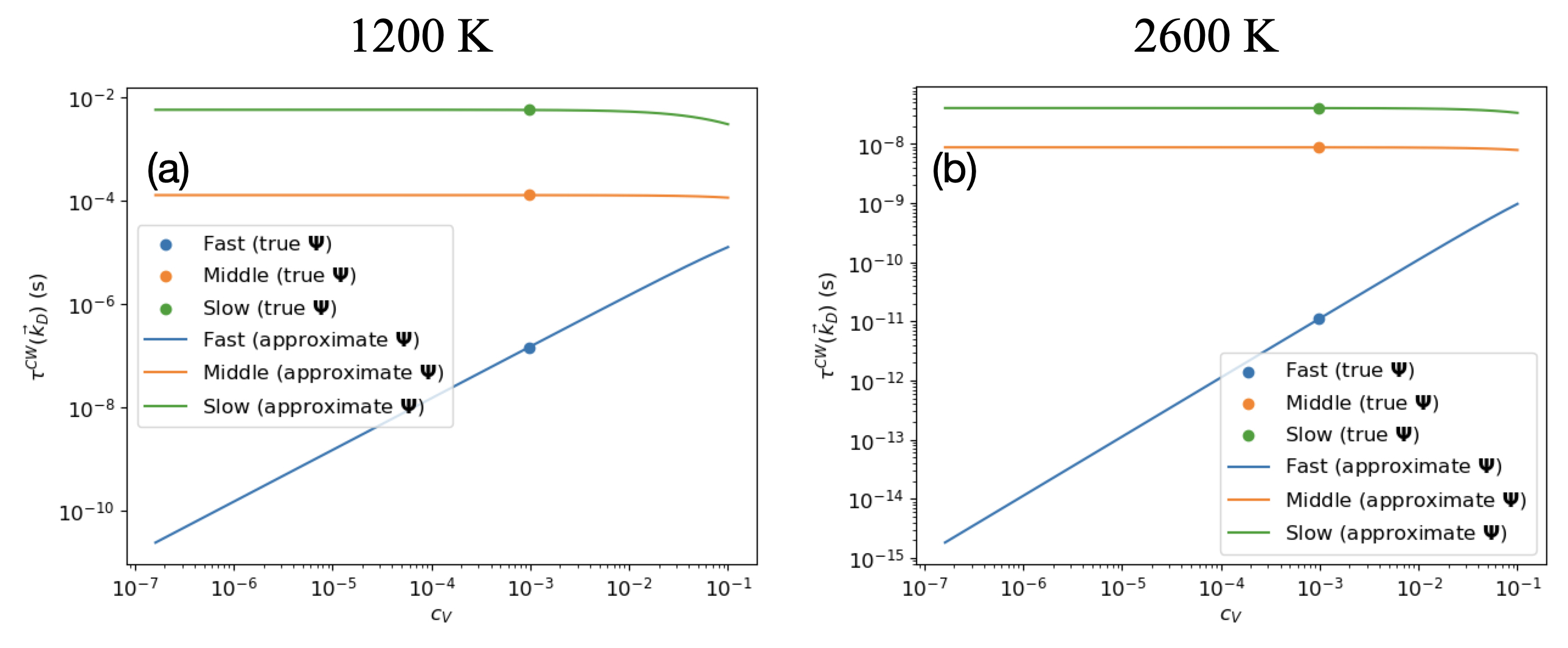}
    \caption{Fast, middle, and slow characteristic relaxation times obtained from the CW model for the $\vec{k}_D$-wave-vector using the true thermodynamic matrix including the vacancy contribution (points) as well as the approximate thermodynamic matrix obtained through treating the vacancy as an ideal species with an arbitrarily dilute concentration (lines) for (a) 1200 K and (b) 2600 K.}
    \label{fig:ideal_vac}
\end{figure}

Relaxation times using the approximate $\bm{\Psi}(\vec{k}_D)$ matrix are presented in Figure \ref{fig:ideal_vac} for temperatures of 1200 K and 2600 K. We compare to results obtained using the true $\bm{\Psi}(\vec{k}_D)$ matrix obtained from long-time kMC simulations including a vacancy. Encouragingly, we see that with the choice of $c_V$ equal to 1/1024, the CW model predictions are nearly on top of each other when using the approximate or true $\bm{\Psi}(\vec{k}_D)$ matrix. We find that the slower relaxation times do not depend on the value of $c_V$ as long as it is below a value of about 0.01, which is a much larger vacancy concentration than would be expected in a real alloy. The fastest relaxation time is strongly dependent on the value of $c_V$. However it is orders of magnitude faster than the other relaxation times, and as evidenced by the timescale of SRO formation from our kMC results, is not relevant to the kinetics of SRO formation. Therefore, our approach to approximate the $\bm{\Psi}(\vec{k}_D)$ matrix with a dilute vacancy concentration does not affect the prediction of $\tau_s(\vec{k}_D)$, and it circumvents the need to parameterize vacancy-atom interactions.

In conclusion, we have shown that the slowest characteristic relaxation time of the dominant ordering wave-vector, $\tau_s(\vec{k}_D)$, predicted by the CW model provides a reliable order of magnitude estimate of isothermal SRO relaxation kinetics in a ternary multicomponent alloy when compared to kMC results. The CW model only needs a mobility matrix and a thermodynamic matrix, which for a real alloy can be obtained from mobility databases and from canonical Monte Carlo simulations using an accurate description of interatomic interactions. We anticipate that this model can aid in designing processing conditions to promote or suppress SRO in multicomponent solid solution alloys.

\section*{Acknowledgements}

A.A. acknowledges support during the initial phases of this work from a fellowship by the UC National Laboratory Fees Research Program of the University of California, Grant Number L21GF3646, and at the National Institute of Standards and Technology through a U.S. National Research Council fellowship. B.X. and P.C. gratefully acknowledge support from the US Department of Energy (DOE), Office of Basic Energy Sciences, under Award No. DE-SC0022295. This work was supported by FUTURE (Fundamental Understanding of Transport Under Reactor Extremes), an Energy Frontier Research Center funded by the U.S. Department of Energy (DOE), Office of Science, Basic Energy Sciences (BES), United States.

\bibliographystyle{elsarticle-num}





% \end{thebibliography}
\end{document}